# How Information Diffuse in Nomination Network?

## - Taking #手写加油接力# Welfare Event on Weibo as an Example

WANG Minghao

XU Keyu

Xiaohui Wang

Paolo Mengoni, IEEE Member

# Abstract

During the special period of the COVID-19 outbreak, this project investigated the driving factors in different information diffusion modes (i.e. broadcasting mode, contagion mode) based on the nomination relations in a social welfare campaign #手写加油接力# on Weibo. Specifically, we mapped a nomination social network and tracked the core communicators in both modes. Besides, we also observed the network from perspectives such as relationships between core communicators and modularity of the whole network. We extracted 6 homophily factors and tested them on 2 representative communities within the largest component of the network. We found that some core communicators distributed in a co-dependent way. At last, we supposed several explanations to the phenomenon which can be explored in further research.

# Introduction

**Background**

Under the influence of the outbreak of COVID-19, Weibo (a Twitter-like social media platform in China) launched a charitable campaign named #手写加油接力# on Feb 3, 2020, aiming to encourage people to relay their blessing to Wuhan, China and to cheer for the workers who are fighting on the front line of the epidemic. The participators of #手写加油接力# Campaign (hereafter referred to as the Campaign) should post a microblog on Weibo with uploading their own handwritten greeting picture, adding the hashtag #手写加油接力# and nominating their friends by @ user ID. Those people who are nominated by others need to pass on this relay. Half a month later, the Campaign had 4.17 billion views and more than 30 million engagements (data on February 18, 2020), which can be understood as a successful viral campaign on Weibo.

**Related works**

The advent of social media has facilitated the study of information diffusion, user interaction and user influence over social networks [1]. In many previous studies on social media and information diffusion, the scholars mainly focused on using retweet relations [3] [4] or follower relations [5] [1] to map and analyze the information diffusion network. Actually, nomination is also a common method to generate social network data [2]. But nomination networks are often used to study the homophily of intimate and small social relationships [7]. The researches on the Ice Bucket Challenge, a campaign gained global recognition from digital audiences as one of the most successful disease-related viral campaigns using social media [8], also focuses on the homophily of celebrities. Few studies have explored the different information diffusion modes (broadcasting and contagion) in the nomination network. Therefore, our project intended to build a social network based on the relationship of the senders and nominees in the Campaign, aiming to explore the driving factors in the different information diffusion modes.

Our research is divided into two main directions: information diffusion mode of this Campaign and the related content of key opinion leaders (also can be regarded as core communicators in this study). For key opinion leaders, we pay attention to the division of different attributes of this group and the structural mode of their collaboration.

**Information diffusion mode on the Weibo platform**

Previous studies have explored the diffusion characteristics of cancer education information on the Weibo platform and found that information diffusion is driven by a mix of the broadcast and contagion mechanisms in the retweet network [3]. At the same time, "Out-degrees" and "in-degrees" can be regarded as the two key structural characteristics of advertisers on social media [9].

Similarly, for this Campaign, we raise the following research question:

RQ1. Whether the information diffusion mode of this Campaign is broadcasting or contagion?

## KOLs and their relationships on the Weibo platform

### *What is KOL*

The definition of key opinion leaders in the sociological phenomenon and social media is different. When we mention this term here, we mainly refer to the latter. That is social media influencer, who has acquired or developed their fame and notability through the Internet [20]. In this group, they can also be subdivided into Micro-celebrities (a person famous within a niche group of users on a social media platform) and Wang Hong (i.e. 网红, the Chinese version of the Internet stardom), etc.

KOLs in this Campaign

During the outbreak of COVID-19 in 2020, online social activities not only be significant in social coordination but also played an important role in the establishment of motivation [10]. On the one hand, the 14-day quarantine period allows people to have more free time to participate in social media. On the other hand, people are more attracted to celebrities, which strengthens their online social status [11]. Therefore, if celebrities are active during the outbreak, it will inspire people to increase the possibility of online socialization.

Meanwhile, the other groups, such as peer leaders, unverified, active, well-connected users and medical professionals with ICT experience, they all can play an active role in promoting SNS (social networking sites) organ donation information [12].

But in this Campaign, who are the promoters? To explore this problem, we have the following question:

RQ2. Who strongly promoted this Campaign?

*The ways and causes of KOLs collaboration in this Campaign*

Homophily refers to "a contact between similar people occurs at a higher rate than among dissimilar people." [13]. It will be easier to form homophonous ties with the same geographic location or common hobbies between members [14]. A study on the online venue for international expats in Denmark shows that the group manifests itself as a community in terms of attachment to geographical location, degree of mutual responsibility of its members, recognition of communal history pieces, and normativity level [15]. In the previous related works on social media homophily, factors like people's political views [16] or ethnicity, religion, age, country, and the reasons for joining specific social media platforms [17] are tested.

It seems to be a common phenomenon that users on social media spontaneously form a community for some reason. Since our research is based on Weibo and the nodes in the network are basically entertainment stars, we decided to choose age, school, occupation, employer, the number of followers, and if they are verified as the factors to analyze the homophily. Then we will discuss:

RQ3. What makes Weibo KOLs form communities of different sizes?

## Method

### Data collection

We used the Python Web Crawler to extract the posts on Weibo that contain the hashtag #手写加油接力#. The popularity of the Campaign has declined after half a month. Therefore, we selected a time period from February 3 to 18, 2020. All information about the posts were collected to our dataset: number of retweets, number of comments, number of likes, content and pictures of posts, posting time and location. The senders' personal profiles were also included: account name and gender, number of followers and followings. In consideration of the complexity of the ordinary accounts, we decided to only extract the verified account. Eventually, we obtained 13740 posts. In the process of data cleaning, firstly, we deleted duplicated rows and then we extracted all nominees from the text of posts that have "@" in it. Because there were many participators without nominating others in their posts and we deleted them. As a result, a total of 4532 posts were retained in the final dataset. Finally, we only retained two columns, the sender name and corresponding nominees, as the source and target node for the visualization stage.

The final dataset was imported into Gephi to generate the visualization of the network. It was a directed graph with 2256 nodes and 4310 unweighted edges.

**Information diffusion mode of the Campaign**

About the broadcasting mode, since the nodes with large out-degree are basically advertisement accounts, so we decided to utilize in-degree together with timeline to examine the broadcasting mode. We first calculated the average in-degree in the network to illustrate the normal scale of diffusion. Then we selected the top 10 nodes with the highest in-degree, and extracted their 1-degree egocentric network. Meanwhile, we checked the time of their activation. So, if a node has a high in-degree and its activation is earlier than its neighbors, we regard this node is of the broadcasting mode.

About the contagion mode, the eccentricity (i.e. the distance from a given starting node to the farthest node from it in the network) of nodes are compared with the diameter of the whole network. If the eccentricity is close to the diameter, we regard this node is of the contagion mode.

**Promoters of the Campaign**

Since we have calculated the key indicators of both modes (i.e. the in-degree and the eccentricity), we regard the top 10 nodes under each indicator are the promoters of the Campaign.

## KOLs' community analysis

We first arranged the nodes into different components using Gephi by their component ID, then we calculated the modularity class applying the algorithm proposed by Blondel et al. [18]. We selected the communities within the largest component and conducted portraits from 6 dimensions (i.e. age, school, occupation, employer, the number of followers and if they are verified) of those users and tried to extract some factors of their homophily.

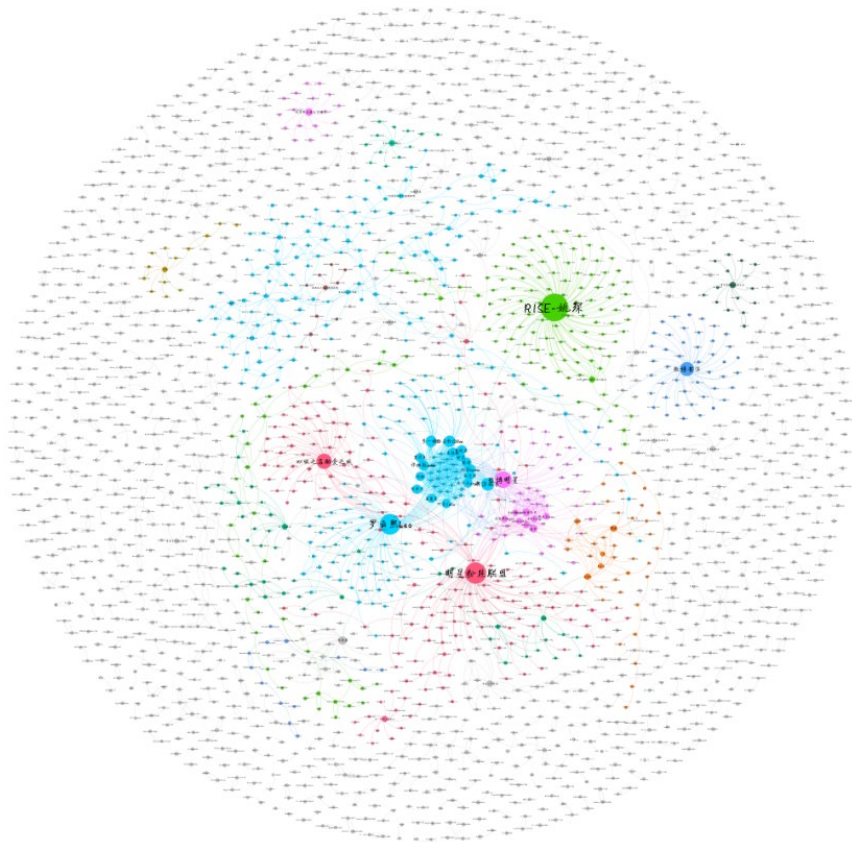

**Figure 1**
*The whole network of the Campaign. The node size corresponds to its in-degree, while different color illustrates different communities.*

**Results**

Figure 1 shows the whole network of the Campaign. It is a directed graph with 2256 nodes and 4310 edges. The network has 551 components and its density is 0.001. The average in-degree of the network is 1.584, and Figure 2 shows the top 10 in-degree nodes and their 1-degree egocentric network (the amount of neighbors are annotated below each chart). These 10 focal nodes have 54.400 nodes connected around them in average, with a standard deviation of 22.916. The timeline rank (i.e. the appearance rank of the focal nodes among the egocentric network in a decreasing time order) is shown in Table 1. As we can see, most focal modes appeared earlier than their neighbors ($M_{timeline\ rank} = 0.175, SD_{timeline\ rank} = 0.312$). So, we can prove there exists the broadcasting mode in the information diffusion of the Campaign.

**Table 1**
*The timeline rank of top 10 in-degree nodes.*

| name | timeline rank |
|---|---|
| R1SE-姚琛 | 0.941 |
| 明星粉丝联盟 | 0.535 |
| 罗云熙Leo | 0.037 |
| 微博明星 | 0.084 |
| 以啵之名助爱之城 | 0.018 |
| 微博书法 | 0.015 |
| 新浪娱乐 | 0.044 |
| 任嘉伦Allen | 0.017 |
| 李一桐Q | 0.022 |
| 谭松韵seven | 0.042 |

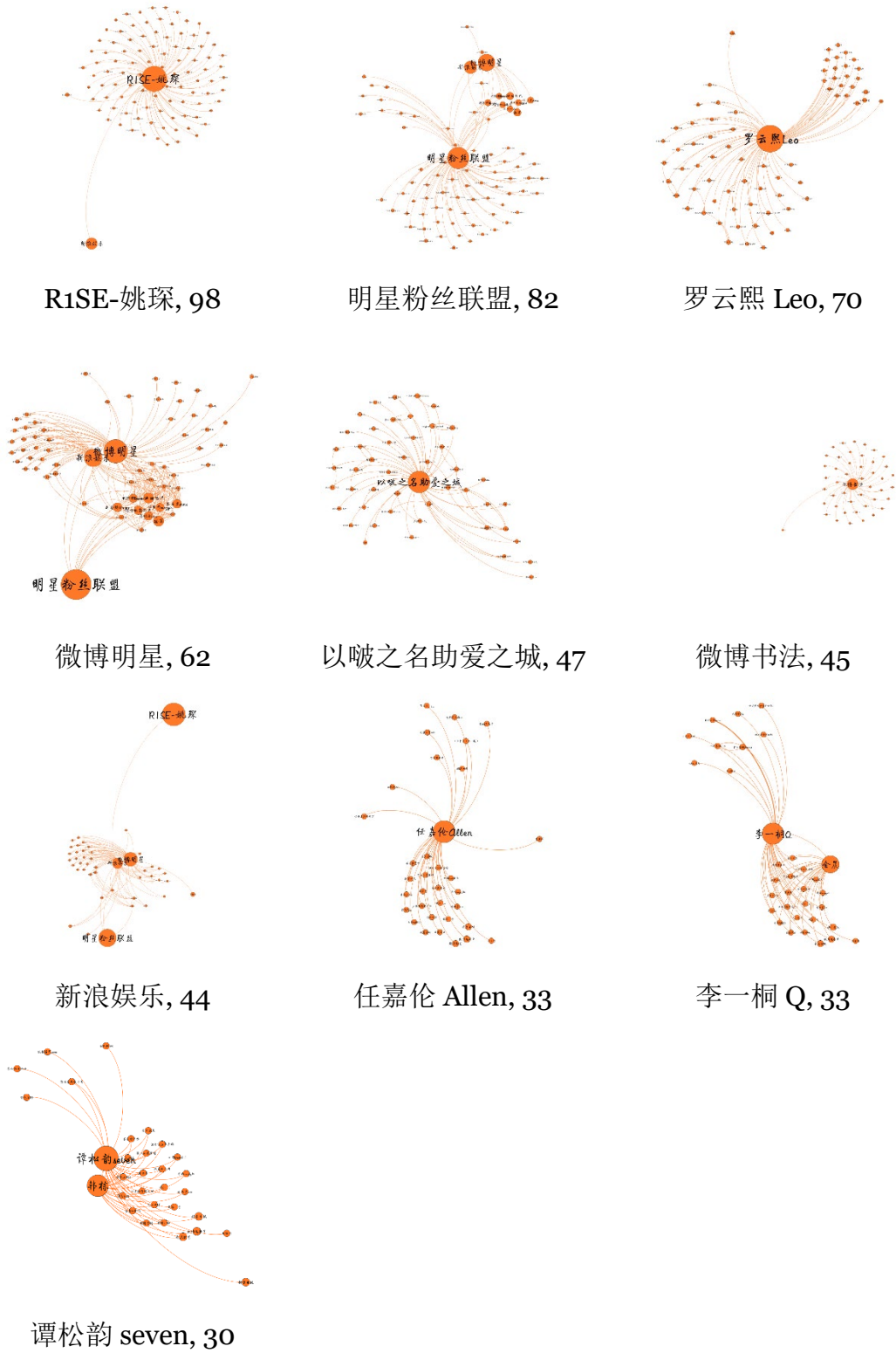

**Figure 2**

Figure 3 shows the longest path in the network. Both the node size and color correspond to the eccentricity (i.e. bigger and darker nodes have larger eccentricity). The top 10 eccentricity are quite similar ($M = 21.800, SD = 0.632$). The diameter of the whole network is 23. So it can be proved that contagion mode is also obvious in the information diffusion of the Campaign.

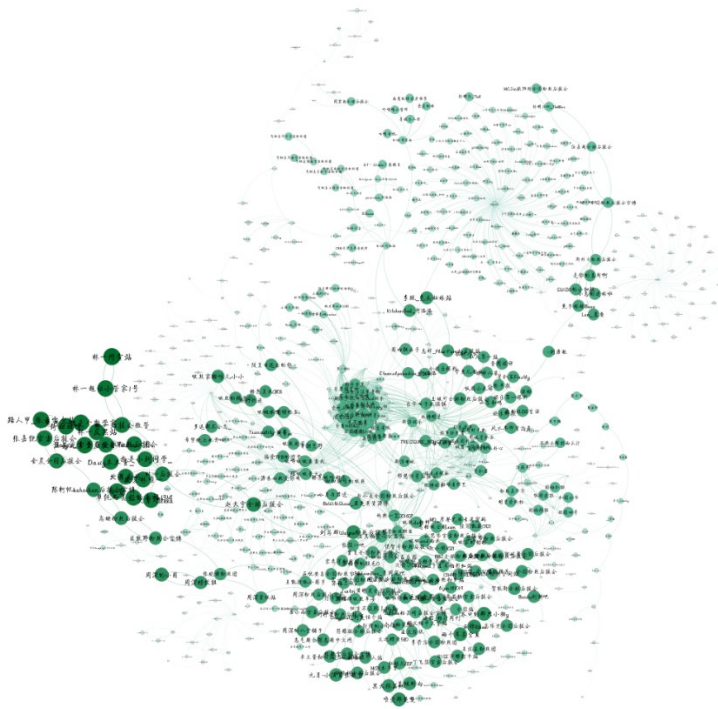

**Figure 3**
*The longest path within the network. Its start node "林一网宣站" has a eccentricity of 23, equals to the network diameter.*

Table 2 illustrates the top promoters of neither mode. There is no overlap between them, and the correlation between them is negative ($r = -0.518$), showing that the relationship between the two modes are fairly negative related. Actually, among the top 10 promoters in the broadcasting mode, a half of them are celebrity themselves, while among the top 10 promoters in the contagion mode, all of them are marketing accounts and fan clubs.

**Table 2**
*The top 10 promoters of neither mode.*

| promoter | mode |
| --- | --- |
| R1SE-姚琛 | broadcasting |
| 明星粉丝联盟 | broadcasting |
| 罗云熙Leo | broadcasting |
| 微博明星 | broadcasting |
| 以啵之名助爱之城 | broadcasting |
| 微博书法 | broadcasting |
| 新浪娱乐 | broadcasting |
| 任嘉伦Allen | broadcasting |
| 李一桐Q | broadcasting |
| 谭松韵seven | broadcasting |
| 林一网宣站 | contagion |
| 路人甲_张云雷个站 | contagion |
| 汪苏泷官方后援会 | contagion |
| 林一超话小管家1号 | contagion |
| 兔子小姐是个哲学家 | contagion |
| 绮妞妞呀 | contagion |
| 林一反黑站 | contagion |
| Boogie_王子异全球粉丝后援会 | contagion |
| 张嘉倪官方后援会 | contagion |
| 我是小刘同学_ | contagion |

Figure 4 shows the communities within the largest component (contains 683 nodes, 30.275% of all nodes) of the whole network. There are mainly 7 communities within this component. We selected two representative communities to conduct portraits.

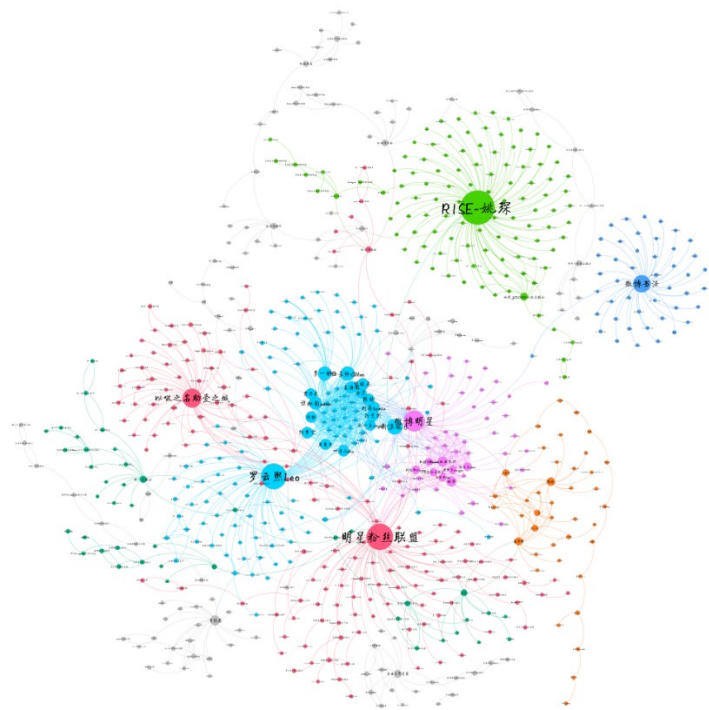

**Figure 4**
*There are mainly 7 communities within the largest component of the network.*

**Discussion**

In the current study, we aimed to discover the information diffusion modes in the topic of ＃手写加油接力＃ on Weibo. By calculating in-degree, we discovered that information disseminates in a broadcasting mode and we also verified that there was a virus-like spreading mode in the network by calculating eccentricity. Then we found that there are different KOLs in two modes. For broadcasting mode, stars were the main group to help to achieve a greater information coverage. For contagion mode, marketing accounts and fan clubs were the KOLs who followed the relay rules to a greater extent and formed deeper connections. To find out how KOLs played a role in the network, we selected several top KOLs to observe their distribution and discovered that they gathered in different communities which could be attributed to homophily.

Besides, we also observed the relationship between different KOLs based on the network. As KOL in the broadcasting mode, stars received in-degree edges from both fan clubs and marketing accounts. Fan club groups always distributed around their idols to support them in the network. Compared to fan clubs, marketing accounts didn't have such significant preference. However, it showed that there were also specific marketing account groups lying around specific star groups in the network. In details, there were two kinds of distributions. The first was one group to one group. One group of marketing accounts was corresponding to one star relay groups. For example, the marketing accounts lying around 杨幂-oriented corporation community only pointed to members in the community. It interpreted that they might be in cooperation or marketing accounts aimed to obtain more attention from the public. The second was one group to many groups. One group of marketing accounts was corresponding to several star relay groups. For example, a fixed group of marketing accounts interacted with both stars from the same crew in a relay chain and stars from other exclusive relay chains. It could be interpreted that marketing accounts might want to obtain more public attention. Another possibility could be it was a marketing method utilized by stars. The phenomenon offered further research insights by analyzing more connections between fixed star groups and marketing accounts in more scenarios.

What's more, we also observed the connection between different communities by the modularity which indicated the dense connection within communities and sparser connections between communities.[19] In our network, modularity was 0.892 which indicated a very high level and showed dense interactions in the event in initial analysis. However, combining indicators of 1848 strongly connected components and 568 communities, we found that a great amount of small size of components were included in the same group by the algorithm [18], which revealed that many users in the event relayed and enjoyed in a limited circle.

## Conclusion

Our research found the KOLs of public welfare relay activities on Weibo. It provides ideas for the development and communication of such activities in the future. According to the communication effect needed to be achieved, find the corresponding KOL, attract more attention and participation, and transfer love and positive energy.

We need to acknowledge there are some limitations in the study. First, we only focused on the influence of one relay event in the public welfare area and there was no comparison between events to draw a general conclusion. Second, we only made analysis on the specific social media of Weibo which may be not applicable in other social medias. More comparison researches between public welfare relay events on different social medias can be a further research direction.

# Appendix

**Table 5**
*Workload contribution*

| Workload (overview)      | Contributor          |
|--------------------------|----------------------|
| research design          | All                  |
| literature study         | LIAO Yihui   XU Keyu |
| data collection          | All                  |
| data cleaning            | WANG Minghao         |
| experiment & visualization | WANG Minghao       |
| report                   | All                  |

| Workload (report)         | Contributor   |
|---------------------------|---------------|
| backgroud & related works | LIAO Yihui    |
| research questions        | XU Keyu       |
| data collection           | LIAO Yihui    |
| experiment                | WANG Minghao  |
| results                   | WANG Minghao  |
| conclusion                | LIU Xiaowen   |
| discussion                | LIU Xiaowen   |
| polish & layout           | WANG Minghao  |